\documentclass[preprint2]{aastex}

\newcommand\plotoneb[1]{
\typeout{Plotone included the file #1}
\centering
\leavevmode
\includegraphics*[width={.85\columnwidth}]{#1}
}

\shorttitle{Masses and Radii of Planet Hosts}
\shortauthors{T.M. Brown}

\begin{document}


\title{Radii of Rapidly-Rotating Stars, with Application to Transiting-Planet 
Hosts}


\author{Timothy M. Brown}
\affil{Las Cumbres Observatory Global Telescope, Goleta, CA 93117}
\email{tbrown@lcogt.net}


\begin{abstract}
The currently favored method for estimating radii and other parameters of
transiting-planet host stars is to match theoretical models to observations
of the stellar mean density $\rho_*$, the effective temperature $T_{\rm eff}$, 
and the composition parameter $[Z]$.
This explicitly model-dependent approach is based on readily-available
observations, and results in small formal errors.
Its performance will be central to the reliability of results from
ground-based transit surveys such as TrES, HAT, and SuperWASP, as well
as to the spaceborne missions MOST, CoRoT, and Kepler.
Here I use two calibration samples of stars 
(eclipsing binaries and stars for which asteroseismic analyses are available)
having well-determined masses and 
radii to estimate the
accuracy and systematic errors inherent in the $\rho_*$ method.
When matching to the Yonsei-Yale stellar evolution models,
I find the most important systematic error results from selection bias 
favoring rapidly-rotating (hence probably magnetically active)
stars among the eclipsing binary sample.
If unaccounted for, this bias leads to a mass-dependent underestimate
of stellar radii by as much as 4\% for stars of 0.4 $M_{\sun}$,
decreasing to zero for masses above about 1.4 $M_{\sun}$.
Relative errors in estimated stellar masses are 3 times larger than those
in radii.
The asteroseismic sample suggests (albeit with significant uncertainty)
that systematic errors 
are small for
slowly-rotating, inactive stars.
Systematic errors arising from failings of the Yonsei-Yale models
of inactive stars
probably exist, but are difficult to assess because of the small number
of well-characterized comparison stars having low mass and slow rotation. 
Poor information about $[Z]$ is an important source of random error,
and may be a minor source of systematic error as well.
With suitable corrections for rotation, it is likely that 
systematic errors in the
$\rho_*$ method can be comparable to or smaller than the random errors,
yielding radii that are accurate to about 2\% for most stars.

\end{abstract}




\section{Motivation}

The radii of transiting extrasolar planets are measured in units of the radius
of the host star.
Thus, to understand the key properties of such planets,
as well as for a sense of their astrophysical context, the masses, radii,
and compositions of the host stars are of great interest.
With the advent of efficient searches for transiting
planets such as TrES, HAT, and SuperWASP \citep{alo04b,bak02,bak04,pol06}, and especially
spaceborne analogs MOST, CoRoT, and Kepler \citep{wal03,bag07,koc04}, the need
for accurate stellar parameter estimates has become acute.
Recently, \citet{soz07}, \citet{bak2007}, \citet{win07}, \citet{cha07}, \citet{tor08} 
and others
have built on work by \citet{mandel02}
to develop the currently-favored approach to stellar parameter estimation,
which for the purposes of this paper I will term ``the $\rho_*$ method''.
This paper aims to understand the precision and accuracy that one may
expect from this method, and, by comparing against other ways of measuring
stellar parameters, to identify the sources and likely
magnitudes of systematic errors that may affect the method's results.

The $\rho_*$ method uses the transit light curve to estimate the 3
parameters $R_p/R_*$, $R_*/a$,
and the transit impact parameter
$b$, which measures the minimum projected distance between the centers
of the star and planet in units of the stellar radius.
Here $R_p$ is the planetary radius,
$R_*$ is the stellar radius, and $a$ is the orbital semimajor axis.
The fitting procedure used to make these estimates relies on a model of
the stellar atmosphere to characterize the stellar limb darkening.
From $R_*/a$, the planet's orbital period $P$ (which ordinarily is known with
extremely high accuracy), and Kepler's third law,
one may compute the stellar mean density $\rho_*$ with an accuracy
that is limited by the photometric precision of the light curve
measurement \citep{sea03}.
This calculation depends only on geometry and on
Newtonian gravitation, so it is independent of modeling assumptions.
Next, by comparing observed high-resolution spectra of the parent star with
model stellar atmospheres, in typical cases
one can estimate $T_{\rm eff}$ with precision of
about 1\%, and $[Z]$, the logarithmic metal abundance relative to solar, 
with quoted errors of about 0.05 dex (e.g. \citet{val05}).
Finally, one searches a grid of stellar evolution models for the best
match, in a $\chi^2$ sense, between computed and observed values 
of $\rho_*$, $T_{\rm eff}$,
and $[Z]$.
The search for optimum model parameters \{mass, age, composition\} is
usually conducted using Markov chain Monte Carlo (MCMC) methods.
These
also yield estimates of the formal distribution of error in
(and if desired, the covariance
between) the various model parameters, and indeed any other global property
of the stellar model ($R_*$, $\rho_*$, $\log(g)$, $T_{\rm eff}$, etc.).

An important virtue of
the $\rho_*$ method is that it typically yields significantly smaller formal 
errors than do older techniques -- uncertainties of 2\% or less in
$R_*$ are commonly achieved.
This is possible because the observables \{$\rho_*$, $T_{\rm eff}$, $[Z]$\}
are individually fairly well-determined, and because each of them
corresponds fairly closely to one of the model parameters
\{$M_*$, $A_*$, $[Z]$\} describing the stellar mass, age, and metallicity.
Thus, during a star's main-sequence lifetime, $T_{\rm eff}$ depends mostly
on $M_*$, $\rho_*$ depends mostly on $A_*$, and (no surprise here)
$[Z]$ depends mostly on $[Z]$.
On the other hand, the method is manifestly model-dependent.
To give correct results, the technique depends upon having 
evolution models that accurately
reproduce the mass-luminosity-radius-composition relations followed by
real stars.
Moreover,  the method depends upon stellar atmospheres models both to account
for limb darkening and to allow interpretation of the stellar spectrum
in terms of $T_{\rm eff}$, $\log(g)$, and $[Z]$. 
There are good reasons to believe that stellar evolution and
atmosphere models in fact
represent real stars fairly well,
but one nevertheless desires direct tests of the $\rho_*$ method.
Fortunately the recent update \citep{tor09} of Andersen's (1991)
classic review 
paper on fundamental parameters of stars in eclipsing binary (EB)
systems collects a large sample of stars with well-determined masses
and radii.
There is also an almost independent but far less numerous group of stars
for which asteroseismology (often combined with other measurements)
provides similarly precise data.
Both of these samples of stars are well enough characterized to
permit tests of the 
behavior of the
$\rho_*$ method.

\citet{tor09} (henceforth TAG) contains the results of analyses of
94 EB systems plus $\alpha$ Cen A and B,
comprising 190 individual stars for which masses and radii are thought
to be known with accuracies of better than 3\%.
The masses and radii thus determined
are almost independent of complex theory, the only exception being the
weak dependence of inferred radii on assumptions about
the stellar limb darkening.
Also given by TAG are estimates of $T_{\rm eff}$
and, for 21 of the systems, $[Z]$.
For most of these systems $T_{\rm eff}$ and $[Z]$ were determined from
photometry, using methods that are not consistent across the sample.
The spectroscopically-derived estimates of these parameters that are
common in the exoplanet context are difficult to obtain for EBs,
because of the blending, broadening, and relative Doppler shifts
that characterize double-line spectra.

From the above information it is straightforward to synthesize the input data
($\rho_*$, $T_{\rm eff}$, and if only by assumption, $[Z]$)
that would be measured if a transiting planet were to orbit any
of the stars.
One may then apply the $\rho_*$ method to each EB component, and compare
the masses and radii that emerge from the model-fitting process to those
that were actually measured.

The stars with asteroseismic data consist almost entirely of ones with
roughly solar mass, and largely of post-main-sequence objects having greater
than solar luminosity.
The observational demands for asteroseismology are severe, so in most cases
the only reliable observable is the so-called ``large frequency separation''
$\Delta \nu$,
which itself depends mostly upon $\rho_*$ \citep{han04}.
In order to obtain a well-constrained result, the measured oscillation
frequencies are often augmented with other kinds of data, such as
interferometric estimates of radii, or dynamical mass estimates in the
case of binary systems.
Finally, detailed interpretation of this information always involves
comparisons between stellar evolution and oscillation models and the
observations.
Thus, the asteroseismically-determined stellar parameters are likely
no more accurate and are certainly as model-dependent as those obtained
from the $\rho_*$ method.
They usually depend on different model implementations, however, so at
worst they tell us something about consistency among extant models of
stellar evolution.
This in turn provides some insight into the poorly-understood 
possibilities for systematic error in the asteroseismic analysis.

I implemented the $\rho_*$ method, using the Yonsei-Yale evolution
tracks \citep{yi01,kim02,yi03,dem04} as the needed stellar evolution models,
and I then applied it to the TAG tabulation of EBs
and to 15 stars with asteroseismic measurements.
In the rest of this paper, I present the computational methods
I used and the results I obtained. 
Section 2 briefly describes methodology and algorithms.
Section 3 gives the results of the comparison, illustrated in
various ways.
I find that the $\rho_*$ method applied to EB components generates
errors in both mass and radius that are small but significant,
and that depend systematically on the basic stellar parameters.
Finally, in Section 4 I investigate the origin of these systematic errors.

\section{Parameter Estimation Methodology}

Estimating stellar parameters first 
requires the observations that are to be
fitted, and their uncertainties.
I took the observables to be $\rho_*$, $T_{\rm eff}$, and $[Z]$.
I derived uncertainties for the latter two quantities in the obvious way,
from the stated uncertainties in the photometric data.
The uncertainty in $\rho_*$ required a different treatment, however.
My chief purpose is 
to assess the systematic errors committed by the $\rho_*$
method because of failings in modeling stellar structure.
For this purpose it makes sense to treat the EB 
and asteroseismic masses and radii as error-free,
and then assign an uncertainty to $\rho_*$ that is small enough that its
effect on the derived parameters is unimportant.
Thus, I took the uncertainty in $\rho_*$ to be about 4.5\%, which is typical of
what might be expected from a very good ground-based
observation based on several transits, or a rather poor spaceborne one.

The central feature of the $\rho_*$ method is the stellar evolution model.
I used the Yonsei-Yale (henceforth YY) evolution tracks downloaded from
their "Evolutionary Tracks" site.
\footnote{http://www.astro.yale.edu/demarque/yystar.html}
These models span masses in the range 0.4 $M_{\sun}$ to
5 $M_{\sun}$, and metallicity between 0.001 and 0.08 ($[Z]$ between -1.230
and +0.673, assuming a solar metallicity of 0.017).
They also employ an initial helium abundance $Y = 0.23 + 2Z$, and a constant
mixing length of 1.7432 times the pressure scale height.
The temporal evolution of each model is tabulated on a nonuniform grid
of stellar age $A_*$, with the same number
of steps for each stellar mass, and a given step number corresponding to the
same evolutionary state (meaning core helium abundance or core mass) for
every stellar mass.
This grid based on evolutionary state is convenient for interpolation
between different stellar masses, since one need not do a search in the
age dimension to locate models of different masses that have otherwise
similar properties.
The YY model grid includes models for which the abundances of high-alpha
elements (O, Ne, Mg, etc.) are enhanced relative to 
the solar composition.
In this study I have not investigated the influence of this degree of
freedom, however.

To apply the YY model grid, I first used the interpolation routines supplied
with the models to resample them onto a uniform grid in $\log M_*$ and
$[Z]$;
for the computations described here I used 150 mass steps between 0.4
$M_{\sun}$ and 5 $M_{\sun}$, and 25 steps in $[Z]$ between -1.230 and
0.673.
I did not resample the tables in the age dimension.
From the resampled grids giving $M_*$, $R_*$, and the luminosity $L_*$,
I then precomputed similar tables giving other quantities of interest,
such as $\rho_*$ and $T_{\rm eff}$.
Finally, to obtain parameters at arbitrary values of \{$A_*$, $M_*$, $[Z]$\},
I interpolated into these resampled grids, using cubic interpolation
in the age-mass plane and linear interpolation in $[Z]$.

The MCMC algorithm has been described by \citet{teg04}, and in the context of
transiting planets, by e.g. \citet{win07} and \citet{cha07}.
I perform the Markov chain random walk in the space of indices into
the 3-dimensional grid of stellar evolution models.
I use a Metropolis-Hastings algorithm with Gibbs sampling and random
permutations of the order in which the indices for age, mass, and $[Z]$
are varied.
The probability distribution for accepting a step that decreases the merit
function was Gaussian, with standard deviations along the 3 axes chosen so
that steps along each axis succeed about 25\% of the time.
In addition to estimates of the mean values and standard deviations of the
model parameters, the software produced many diagnostics.
These included marginal probability distributions for every parameter
of interest, scatter plots, and various convergence measures, so that
misbehavior of the MCMC algorithm or ambiguities in the models
could be identified.
I found the most useful convergence test was visual inspection of the
parameter chains, the corresponding chain of $\chi^2$ values, and various
2-dimensional scatter plots of these quantities.
In doubtful cases I reran the MCMC process using chains that were lengthened
by factors of 2 and 4;  these longer runs always yielded chains that were
consistent with each other, and that gave convincingly smooth and well-sampled
sample distributions in parameter space.  

The merit function itself was intended to represent the posterior probability
attached to each set of stellar model parameters, taking into account the
observations.
From a Bayesian perspective, the logarithm of this probability is the sum
of $-\chi^2/2$, representing the probability that the model is
consistent with the observations, and the logarithm of a prior probability,
representing the distribution of model probabilities in the absence of
any observational data.
Each of the parameters \{$A_*$, $M_*$, $[Z]$\} has a prior probability
associated with it, but these are justified in different ways for the
various parameters.

An implication of the time grid chosen for the models was that the interval
(in years) corresponding to a single time step was highly variable, both
for a single choice of mass and composition, and from one such choice to
another.
The variations in this probability are artifacts of the model implementation
and can be quite large,
so it seemed important to account for them.
Therefore I took the prior probability for each model at each age to
be proportional to the duration of the corresponding time step, normalized
by the maximum age of the star represented by the model.
Results of the MCMC process seldom depended strongly on whether I imposed
the age-dependent prior probability.
Exceptions to this rule occurred for stars having similar radii and
temperatures at two distinct ages, one before and one after arrival on
the main sequence.
In these cases the age prior probability tended to discriminate against
the pre-main-sequence phase, because it is very short-lived.

One could also impose prior probabilities based on the statistical distribution
of stars (by mass and by composition) in the solar neighborhood.
I elected not to do this, since the detailed choices of priors in these
cases are somewhat arbitrary, and in any case the observed values of
mass and $[Z]$ are so tightly constrained that reasonable prior probabilities  
have little effect on the calculation's results.
 
Although the MCMC procedure gave a robust estimate of mean values
and corresponding dispersions, it was an inefficient way to determine the exact
position of the merit function optimum.
But this information was valuable for estimating sensitivities
(eg to assumed biases in $T_{\rm eff}$ or $[Z]$), which must be calculated
by measuring the change in optimum values for small changes in the
observations.
For this reason I also computed the parameters giving the optimum merit 
function using an ``amoeba'' search, which when started near the best
parameters found by MCMC,
converged to the local optimum with negligible error in a few tens 
of iterations.
In what follows, unless otherwise stated, 
I use the amoeba-search values for estimated parameters,
with uncertainties computed from the MCMC results.

The probability distributions generated by the MCMC method
were usually close to multivariate Gaussian, 
reflecting the assumed distribution of
errors in the input data.
But in a substantial minority of cases, the stellar models permitted multiple
solutions that fit the data almost equally well.
In such cases I encountered the usual problems associated with
complicated merit-function behavior,
among them that mean values of parameters found by MCMC did not agree
with the optimum values found by the amoeba search, and the amoeba
process itself was likely to yield different results
depending upon its starting values.
These problems add scatter to the results (particularly when
computing sensitivities of stellar parameters to changing input data),
but these uncertainties are not large enough to affect the conclusions.

\section{Application to Eclipsing Binaries}

The list of stars supplied by TAG contained 95 binaries, or 190 stars
for which masses, radii, and uncertainty estimates were provided.
Of these, 34 had $M_* \geq 5M_{\sun}$ or $M \leq 0.4 M_{\sun}$, which are the
limits of the mass range covered by the YY models.
This left 156 stars that could be analyzed by the $\rho_*$ procedure.
Only 32 of these had individual estimates of $[Fe/H]$,
which I simply equated to $[Z]$.
For a few stars, the estimated $[Z]$ values were derived not from
direct observation of the star in question, but from the star's
presumed parent population (star cluster or galaxy).
For stars without explicit measurements of metallicity, I assigned
$[Z]=0.0$, with an RMS uncertainty of 0.2 dex.
These parameters fairly accurately characterize the distribution of EBs
having observational estimates of $[Z]$, though the mean metallicity of
stars in the solar neighborhood is smaller than this by about 0.17 dex
\citep{nor04}.
Effective temperatures for the TAG stars came almost entirely from 
photometry;
for stars cooler than 8000 K
the estimated uncertainties in $T_{\rm eff}$ were typically between
50 $K$ to 250 $K$, though for some hotter stars the uncertainties
were as large as 800 K.
Data for the 156 stars used in this analysis are given in Table 1.

For each star, I applied the MCMC procedure with the search over stellar
models constrained by the star's mean density (computed from the TAG mass
and radius), $T_{\rm eff}$, and $[Z]$.
The uncertainties I assumed in these quantities are described
above.
The MCMC/amoeba procedure then yielded new estimates of 
the stellar mass, radius,
luminosity, age, mean density, $T_{\rm eff}$, and $[Z]$, where the procedure
necessarily returned values of the last
3 quantities that
were much less than 1 standard 

\begin{figure}
\plotoneb{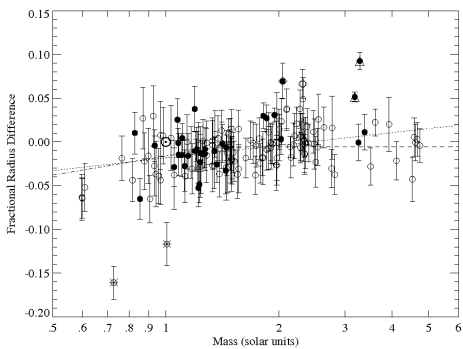}
\caption{
The relative radius discrepancy $\delta R_*$ for stars in the TAG
sample, plotted against stellar mass $M_*$.
The error bars show 1-$\sigma$ (assumed symmetrical) errors derived from
the MCMC analysis, added in quadrature to the radius errors quoted
by TAG.
The stars for which there are individual estimates of $[Z]$ are shown as
filled circles.
I show the Sun's position as a large $\sun$ symbol.
the dotted line show the results of a fit to
$\delta R_*$ as a linear function of $\log M_*$.
The dashed line is a fit to the piecewise linear function of $\log M_*$
described in the text.
Stars excluded from these fits are plotted with special symbols,
and are described in the text.
\label{fig1}}
\end{figure}

deviation from those that were
provided as input.
The values of $M_*$, $R_*$, and $A_*$ derived by this method are
listed in the last 3 columns (labeled ``fit'') 
of Table 1, and in the corresponding
columns of Table 2.
 
To compare input and output masses and radii, for each star I computed
the normalized discrepancy
$\delta M_* \ = (M_{output} - M_{input})/M_{input}$,
and $\delta R_* = (R_{output} - R_{input})/R_{input}$.
Figure 1 shows $\delta R_*$ plotted against
the stellar masses estimated by TAG.
Since the fits included a strong constraint on the stellar mean density,
(as calculated from the provided values of $M_*$ and $R_*$)
I obtained the nearly perfect equality $\delta M_* \ = \ 
3 \delta R_*$.
One of the quantities $\delta M_*$ or $\delta R_*$ is therefore redundant,
and can easily be computed from the other.
In what follows, I will frame the discussion in terms of the discrepancy
in radius $\delta R_*$.

From the Figure, it is evident that the $\rho_*$ method of
estimating stellar radii works fairly well.
Aside from a few outliers, the RMS deviation of $\delta R_*$ is 2.6\%,
while that of $\delta M_*$ (not plotted) is 3 times as large at 7.8\%.
The discrepancies are, however, predominantly negative,
and they vary systematically with stellar
mass.
The sense of this trend is that the masses and radii derived using the
$\rho_*$ method are generally too small, and in percentage terms, the more so
for smaller stars.
I have plotted discrepancies as a function of the stellar mass because
this parameter is independent of stellar age.
But since almost all of the stars on this plot are on the main-sequence or only
slightly evolved, changing the independent variable to stellar radius
or $T_{\rm eff}$ would show essentially the same picture.
These parameters are so tightly correlated along this part of the main sequence
that without further information,
one can hardly say which of them
(or what combination of them) is most central to the observed effect.

Fitting $\delta R_*$ to a linear function of $\log M_*$ yields
$$
\delta R_*  \ = \ (-0.0175 \pm 0.0023)\  + \  (0.0561 \pm 0.0076) \log M_*
\eqno(1)
$$
This fit shows unambiguously that $\delta R_*$ is negative and has a trend
with mass.
The coefficients are nonzero with high statistical significance -- more than
6$\sigma$ for each, with the fits removing about one-third of the variance
in the data set, and shrinking the $\chi^2$ statistic by 40.
Of course, one can also fit more complicated functions to the data.
For instance, a piecewise linear function that has zero slope for
$M_* \ \geq \ 1.4 M_{\sun}$ fits the data marginally (but not compellingly)
worse than does a
single linear function;  this is shown as dashed lines in the Figure.
It is described by
$$
\delta R_* \ = \ (0.0980 \pm 0.020) \log(M_*/1.4), \ M_* < 1.4 \eqno(2)
$$
$$
\delta R_* \ = \ (-0.0021 \pm .098), \ M_* \geq 1.4
$$ 

I excluded 6 outlying stars from the least-squares fits in Fig. 1.
These are the two binaries V1174 Ori and OGLE 051019, as well as
the primary components of SZ Cen and TZ For.
All of these stars except perhaps SZ Cen are unusual as regards their
evolutionary state or metallicity.
Their individual peculiarities are described in 
section 4.4 below.

So far I have discussed only the masses and radii of stars in the TAG
sample.
Considering the best-fitting age values gives further evidence 
that the evolution models are failing for at least some stars.
Stars with masses larger than 1 $M_{\sun}$ have, almost without exception,
inferred ages that are less (usually much less) than that of the Milky Way.
On the other hand, stars below solar mass have, with very few exceptions,
inferred ages considerably greater than the Sun's, and most often
greater than a Hubble time.
The record-holder in this regard is V1174 Ori B, with an apparent age
that is greater than 100 GY, but many other low-mass stars display
impossible ages.

The likely reason for these excessive age estimates is that
the low-mass stars in the TAG sample have mean densities that are so low
that stars with the observed values of $T_{\rm eff}$ and $[Z]$ attain them
only after they are significantly evolved.
Given the extended main-sequence lifetimes of these stars,
such evolution can take a very long time.
A related possibility is that, for low-mass stars, the measured
effective temperatures are systematically too low.
If this were the case, then the ZAMS model for a cool star would,
in addition to being too cool, be smaller and more dense than it should be. 
Getting such a model to the observed $\rho_*$ would then require
greater age than it should, and perhaps greater than possible.
In either case, one can be sure that model-fitting solutions that
require unphysical ages indicate something wrong with the models.

\section{Discussion}

Systematic differences between YY models and the samples of
stars described in the previous section
are concerning, even if they are rather small.
But the way in which the differences are interpreted (and perhaps
corrected) depends on how they arise.
Three general sources of systematic error seem possible.
First, the masses and radii estimated by the $\rho_*$ method may be
incorrect because of systematic errors in the input data.
Second, the chosen samples of stars may be systematically different from
the galactic mean, or at least from those stars that one expects to 
encounter in planet searches.
Last, the evolution models may simply be erroneous in some respects.
Any or all of these possibilities may be true to some extent.
I consider each of them below.

\subsection{Systematic Errors in Input Data}

Biased input data may of course lead to biased estimates of stellar 
mass and radius.
In particular, values of $T_{\rm eff}$ and $[Z]$ derived from comparisons
between observed spectra and stellar atmospheres models may not correspond
exactly to $T_{\rm eff}$ and $[Z]$ as used in the stellar evolution models.
Moreover, such inaccurate correspondence might plausibly depend on
$T_{\rm eff}$ itself, which in principle could generate the observed lack
of agreement.
($\rho_*$ might be in error also, but as applied to the
TAG sample of EBs and to the seismically-constrained stars
it is correct by definition,
and as applied to transiting planets it depends only on such well-established
physics that its reliability seems assured.)

Systematically inaccurate reddening estimates might also influence the
masses and radii, by generating systematic errors in the observed
$T_{\rm eff}$ values.
This effect would however be strongest for the hottest and most luminous
stars in the sample, since such stars tend to be the most reddened.
There is no evidence in Fig. 1 for an effect of this sort, so I have
not pursued this possibility.

To assess whether biased $T_{\rm eff}$ or $[Z]$ values might account for the
observed discrepancy, I
estimated the sensitivities
$\partial \log R_*/\partial \log T_{\rm eff}$ and $\partial \log R_*/\partial [Z]$ by
perturbing the input values for the TAG sample.
Figure 2 shows  these sensitivities 
as a function of $\log M_*$,
computed using perturbations of 0.007 in $\log T_{\rm eff}$ and 0.1 dex in $[Z]$.

\begin{figure}
\plotoneb{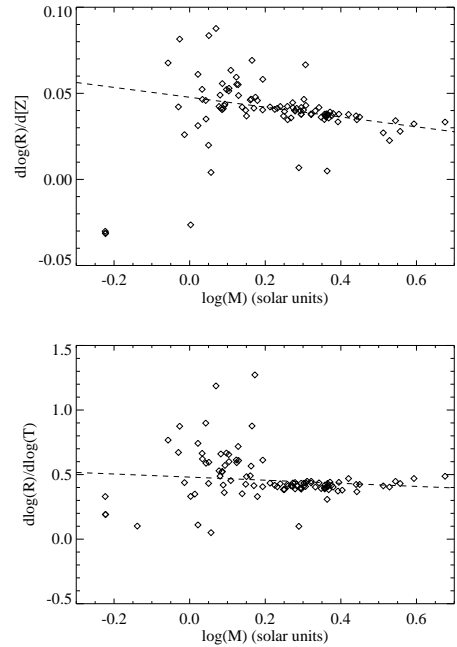}
\caption{Sensitivities of the estimate of $R_*$ resulting from the $\rho_*$
method for stars in the TAG sample of eclipsing binaries. 
The sensitivity to $[Z]$ (upper panel) was computed using $\delta [Z] = 0.1$;
the sensitivity to $T_{\rm eff}$ (lower panel) used $\delta T_{\rm eff} = 0.007$.
Dashed lines show the results of robust fits to linear functions of
$\log M_*$.
\label{fig2}}
\end{figure}

\begin{figure}
\plotoneb{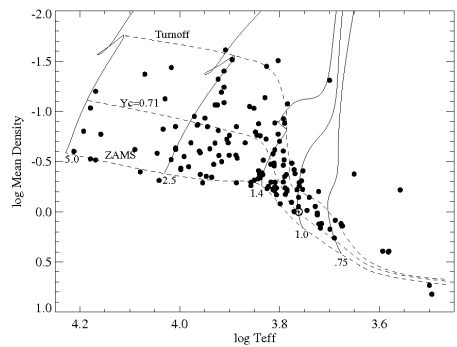}
\caption{$\rho_* \ vs \ \log T_{\rm eff}$ plot for the stars in the TAG sample.
Yonsei-Yale
evolutionary tracks for stars with $M_* =$ \{0.75, 1., 1.4, 2., 5.\}$M_{\sun}$
are overplotted as solid lines.
Loci of constant central $Y$ are plotted as dashed lines, corresponding to
the ZAMS, $Y_c$(solar), and main-sequence turnoff.
The Sun is indicated with a $\sun$ symbol.
\label{fig3}}
\end{figure}

Not all of the binaries yielded valid differences, because of jumps
between multiple solutions as described in section 2.
For the majority of stars, however, the results indicate that the
sensitivities are only weakly mass-dependent. 
A robust straight-line fit to the sensitivities yields
$$
\partial \log R_* / \partial \log T_{\rm eff} \ = \ 0.48 \ - \ 0.12 \log M_* \ 
\eqno(3)
$$
$$
\partial \log R_* / \partial [Z] \ = \ 0.048 \ - \ 0.029 \log M_* \ .
\eqno(4)
$$
Both of these sensitivities are well defined for $\log M_* \geq 0.15$,
and less so for smaller masses.
Even at 1 solar mass, however, it seems safe to estimate that
to cause the
observed -1.7\% discrepancy in $\delta R_*$ would require systematic errors of
about 200 K at $T_{\rm eff} = 5800$ K, or more than 0.15 dex in $[Z]$.
Systematic errors of this size are essentially impossible for $T_{\rm eff}$,
and implausible for $[Z]$.
Thus, while one cannot altogether discount systematic errors in these
observed parameters as contributing to the discrepancy between TAG parameters
and those from the $\rho_*$ analysis, it seems unlikely that they
make an important contribution.

\subsection{Bias in the TAG Sample}

Perhaps the most striking point to be made from Fig. 1 is that the Sun,
to which the YY models are calibrated, is not closely representative of
the stars of its general type in this sample.
Compared to its neighbors in Fig. 1, it is assigned a larger mass and radius
than is typical, by about 5\% and 1.6\%, respectively.
These discrepancies clearly do not arise from errors in the Sun's properties.
There seem to be three possibilities:
the EB stars that fall near the Sun in mass and radius are
subject to systematic errors of a few percent in their mass and radius
measurements, or they are subject to selection bias of similar magnitude,
or there is genuine star-to-star variability in the
stellar mass-radius-luminosity relationship.
Again, none of these possibilities excludes the others;  all may
be acting to some degree.

To address these issues it is helpful to display the data in a way
that separates the observational data (EB masses and radii) from
issues of evolution modeling and data fitting.
Figure 3 shows $\rho_*$ plotted against $\log T_{\rm eff}$ for the TAG
sample of stars, and for the Sun.
The axes are oriented as in a color-temperature diagram, with temperature
increasing to the left, and $\rho_*^{-1}$ (a proxy for luminosity)
increasing upwards.
Overlaid are (solid lines) YY evolution tracks for a sampling of stellar masses,
and (dashed lines) loci corresponding to the zero-age main sequence (ZAMS),
main-sequence middle age (with about 71\% helium in
the core, corresponding to the Sun's evolutionary state), 
and the age of turnoff from the main sequence.
Even more than in Fig. 1, the Sun lies at or near one boundary of the region populated
by stars:  it has $\rho_*$ as large or larger than all
other stars with similar $T_{\rm eff}$.
Evolution in stars of roughly solar mass carries them upward on
this diagram, to lower $\rho_*$ and higher luminosity.
So, barring a statistical fluke, it seems either that
all of the low-mass stars in the TAG sample are at least as
evolved as the Sun, or that evolution models that accurately describes the Sun
do not describe the average properties of the TAG sample.
 
Two obvious sources of bias in the TAG sample
 are Malmquist bias, which will favor detection of
more luminous binaries, and a geometrical bias towards stars with larger radii,
since (assuming random orbital inclinations) these are more likely to
present themselves as EBs.
These effects are however inadequate to account for the strength or the
morphology of the observed bias.
According to the YY evolution tracks, the present-day Sun is about 36\%
more luminous and 12\% larger in radius than it was on the ZAMS.
Its detectability as the larger member of an EB should be proportional
to the volume within which it is brighter than some limiting apparent
magnitude, multiplied by the ratio of solid angles within which the
rotation axis must lie in order for eclipses to occur.
The first factor is at most proportional to $L_*^{3/2}$, while for
well-detached EBs such as found in the TAG sample, the second factor is
proportional to $R_*$.
Sun-like stars in the second half of their main-sequence lifetime thus should
be over-represented among the TAG stars by a factor of at most about 1.8,
relative to younger stars of similar mass.
It is surprising to find the latter group totally absent. 
Moreover, the factors favoring detection of somewhat evolved stars should
be similar for stars of all masses, whereas the trend seen in Fig. 3
is quite different.
For stars with masses greater than 1.4 $M_{\sun}$, the ZAMS is
coincident with the observed maximum-$\rho_*$ boundary, and 
for more massive stars there are roughly equal
numbers in the first and last halves of their main-sequence
lifetimes.
But for masses below 0.8 $M_{\sun}$, no stars are found anywhere 
within the YY main sequence
(although the significance of this observation is shaky because of the small
number of TAG stars in this mass range).  
On balance, it seems unlikely that Malmquist and geometrical selection
bias by themselves provide an adequate explanation for the paucity 
of high-$\rho_*$
Sun-like stars in the TAG sample.

Since most of the TAG stars lack metallicity determinations, another
possible bias is that the true $[Z]$ values for these stars differs
systematically from the assumed Gaussian distribution with $[Z]=0.0\pm 0.2$.
To test this idea I performed fits to the observed $\delta R$ vs. $\log M_*$
dependence, using only the 32 stars with measured metallicities.
For both the linear and piecewise linear fits, the coefficients of the
mass-dependent part of the variation agreed with those in
Eqs. (1-2) within 1$\sigma$, and
for both fits these coefficients differed from zero by about 2.5$\sigma$.
Thus, the subset of stars with measured metallicities gives results that
are consistent with those from the full sample, but with worse precision.

To explore a possible $[Z]$ bias further, I artificially altered the
model $[Z]$ values as a linear function of $\log T_{\rm eff}$, in order
to remove the mass dependence seen in Fig 1.
For all but the least massive ($M \leq 0.6 M_{\sun}$) stars, I achieved
this by increasing $[Z]$ by an amount $\delta [Z] = +0.1 - 
2 \log (T_{\rm eff}/T_{\sun})$, but only for $T_{\rm eff} \leq 6400$K.
This implies that to produce the observed radii, the $\simeq$70 low-mass
stars in the TAG sample would need to be systematically metal-rich
compared to the Sun; about 25\% would require $[Z] \geq 0.2$, and about
10\% would need $[Z] \geq 0.4$.
While such a distribution of $[Z]$ is conceivable, it is implausible
{\it a priori}.
A survey of about 12,000 bright stars in the solar neighborhood
\citep{nor04} shows a distribution that is centered at $[Z] \simeq -0.17$,
with a characteristic width of about 0.22 dex.
Significantly metal-rich stars are therefore fairly rare;
only 4\% of Nordstr\"om's sample show $[Z] \geq 0.2$, and fewer than 0.4\% have
$[Z] \geq 0.4$.
Since no obvious metallicity selection is operating in the choice of
EBs in the TAG sample, it is implausible that the sample should be
as skewed towards high $[Z]$ as is required to explain Fig. 1.
On the other hand, systematically high $[Z]$ would help reduce the large
ages attributed to many low-mass stars,
because increasing metallicity tends to decrease stellar mean density at
constant $T_{\rm eff}$ and age.
A final curiosity is that the very lowest-mass stars (smaller than about
0.6 $M_{\sun}$) show radius changes with $[Z]$ that are smaller than
(and sometimes opposite in sign to) those of their more massive brethren.
Evidently it is dangerous to draw conclusions base on these stars,
presumably because there are many changes in important physical
processes as one nears the boundary between K and M stars.

A bias that is more likely to be important relates to 
the stars' angular momenta and magnetic activity.
\citet{tor06}, \citet{lop07}, and \citet{mor08} have noted 
that some rapidly-rotating, magnetically-active
EB component stars have temperatures that are lower and radii that
are larger than expected for their masses or luminosities.
Conspicuous examples of such stars include 
V1061 Cyg \citep{tor06}, and CV Boo \citep{tor09}.
Both models \citep{chab07,cla09} and general structural considerations 
suggest that
magnetic processes in stellar outer convection zones are responsible
for inflating stars in this way.
Moreover, as pointed out by TAG and others, 
EB systems are strongly selected for small, short-period orbits.
As a result, their components usually are tidally locked, and rotate
rapidly compared to all but the youngest field
stars.
Thus, one expects the TAG sample to be biased in the sense that its
members have larger radii and smaller $T_{\rm eff}$ than would a similar
sample of single stars.
The paucity of high-density of stars 
seen in Fig. 3 is consistent with this expectation.
The lack of dense stars is most evident for sub-solar masses 
(for which stars have deep
convection zones), and weak or absent  for stars above 1.4 $M_{\sun}$,
for which surface convection zones almost vanish.

\subsection{Asteroseismic Sample}

The 15 stars with properties listed in Table 2 have measurements of
their pulsation frequencies that constrain their large frequency
separations $\Delta \nu$, and hence the values of $\rho_*$
(e.g., \citet{han04}).
When combined with other astronomical data (the details vary from
star to star), this information allows accurate estimates of $R_*$ and
$M_*$.
In most cases, the stellar oscillations have been observed in the
radial velocity signal;
for this to be feasible, the stars must have small values of the
rotational line broadening $v \sin i$.
Hence, these stars should not suffer from the rotation/activity bias
ascribed to the TAG sample.
Two stars, $\alpha$ Cen A and B, appear in both the TAG and the asteroseismic
lists.

Fig. 4 shows the result of
applying the $\rho_*$ method to the asteroseismic sample of stars and
plotting mass and radius discrepancies as in Fig. 1.
Expressed in the same terms as Eqns (1) and (2), a fit to these data yields
$$
\delta R_* \ = \ (0.0013 \pm 0.0068) \ - \ (0.002 \pm 0.046) \log M_*
\eqno(5)
$$
and
$$
\delta R_* \ = \ (0.042 \pm 0.075) \log(M_*/1.4), \ M_* < 1.4
$$
$$
\delta R_* \ = \ (0.005 \pm .009), \ M_* \geq 1.4
\eqno(6)
$$

Evidently the YY models fits the asteroseismology sample within the errors.
Moreover, the zero points of the linear fits to the TAG and asteroseismology 
samples disagree by
2.5 $\sigma$, so there is moderately strong evidence that the two distributions
are indeed drawn from different populations.
The fits also yield discrepant slopes for the two samples, but the
asteroseismology sample slope is so poorly constrained that the difference
between slopes is only about 1 $\sigma$.

Figure 5 shows the $\rho_*$ {\it vs} $T_{\rm eff}$ plot for the asteroseismic sample.
One star, $\zeta$ Hyi, is a true giant with $\log \rho_* = -2.6$,
and hence falls outside the plotted range of the figure.
There are 4 stars out of 15 in the asteroseismology 
group that have equal or greater
density and equal or younger evolutionary age than the Sun.
This is consistent with the hypothesis that low-mass EB component stars
have systematically lower density than do slowly-rotating field stars
of the same $T_{\rm eff}$. 
As yet, however, the number of stars with asteroseismic mass and radius
measurements is too small to justify a
firm conclusion in this regard.
Moreover, as noted earlier, asteroseismic parameter estimation has its
own model dependencies.
Agreement between asteroseismic and $\rho_*$ methods for estimating
parameters may say more about similarities between the models used than
it says about the accuracy of either set of models in representing
real stars.

\begin{figure}
\plotoneb{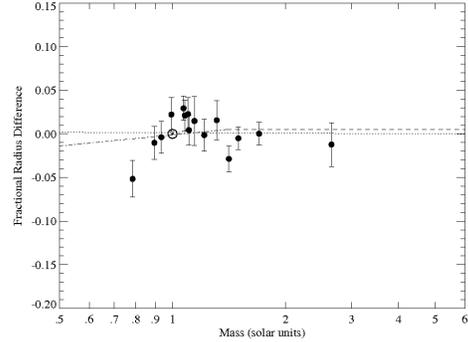}
\caption{Same as Fig. 1, but for stars with asteroseismic estimates of
$\rho_*$.
\label{fig4}}
\end{figure}

\begin{figure}
\plotoneb{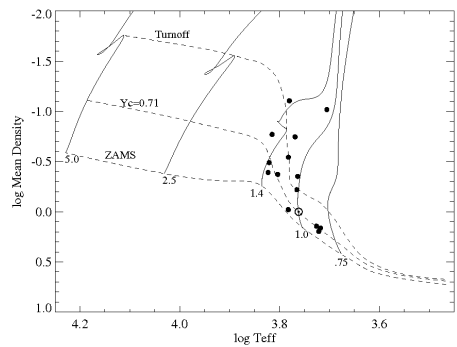}
\caption{Same as Fig. 3, but for the sample of stars with asteroseismic
estimates of $M_*$ and $R_*$.
\label{fig5}}
\end{figure}

\subsection{Stars with Unusual Properties}

In Figures 1 and 4, a few stars stand out as being significantly 
inconsistent with the mean relation between $\delta R_*$ and $M_*$.
I did not include these stars in the least-squares fits described in
section 3, so it is worthwhile to 
understand in what other ways they differ
from the norm. 
In most cases there is a ready explanation for their odd radii.

The two stars overplotted with asterisks in Fig. 2 
have fractional radius differences
smaller than -0.1.
They are the components of VB1174 Ori, a pre-main-sequence
system \citep{sta04} with oversize components that are still contracting.

The two stars overplotted with triangles are components of
OGLE-05019.64-685812.3 \citep{pie09}.
This member of the Large Magellanic Cloud is the faintest and 
the second most metal-deficient
system in the TAG list of binaries (after V432 Aur).
Its assigned metallicity ($[Z] = -0.5$) has not been measured directly;
rather it is inferred from the mean metallicity of its population.
Its small radius relative to the YY models might therefore  result
if its metallicity were actually about $[Z] = -1.1$,
or as a result of errors in the models
themselves, or in how the latter are used.
The last explanation seems most likely, as these stars may in fact be
clump giants, which lie beyond the oldest evolutionary stage
tabulated in the YY models.

The 2 stars with $\delta R_* \simeq +0.05$ and overplotted with 
diamonds are the primary components of
SZ Cen (with $M_* = 2.31$) and TZ For ($M_* = 2.05$).
The SZ Cen system is poorly represented by evolution tracks for coeval
stars with equal initial composition \citep{gro77}, but the reason
for this failure is obscure.
TZ For consists of a pair of giants, with the primary component's radius
being more than 8 $R_{\sun}$ \citep{and91b}.
These authors found that
the more luminous, cooler component of this system (identified as
TZ For A in the compilation by TAG, but as TZ For B in \citet{and91b})
can be reconciled with the properties its companion
only if the former is in the core helium burning stage of
evolution.
Since the YY models extend only to the tip of the giant branch, this
star cannot be represented by them.

Finally, the lowest-mass and most discrepant star in the list
of stars with asteroseismic data is
$\tau$ Cet \citep{tei09}. 
This star is well known for having extremely low magnetic activity,
yet its asteroseismic radius appears 5\% larger than the YY models
would indicate given its $T_{\rm eff}$ and $[Z]$, even if one allows 
the unreasonable age of 24 GY.
In this case the asteroseismic data were compromised by instrumental
noise, and hence may be open to question.
Rather than speculate about causes, it seems best in this instance
to wait for further observations.
\parindent=12pt

\subsection{Conclusions}

The original motivation for this paper was to assess the accuracy of the
$\rho_*$ method for estimating radii of the host stars of transiting extrasolar
planets.
Since the reliability of the method is almost entirely dependent upon
the reliability of stellar evolution models,
this question can also be framed in terms of the ability of such
models (in the present case, the Yonsei-Yale models) to represent
the true mass-radius-luminosity dependencies of real stars.
The short answer to the initial question is yes, 
without any corrections or adjustments the $\rho_*$ method
implemented with the YY evolution tracks will yield stellar radii
that are accurate within (at worst) about 5\%, for a large majority of stars.
This accuracy in radius is already good enough to support useful measurements
of planetary density from transit measurements by Kepler and other
spaceborne photometry missions \citep{sea07}, although the corresponding
15\% uncertainty in mass may be problematic.
However, if one confines attention to the asteroseismic sample of single,
slowly-rotating stars (Table 2), the accuracy of $\rho_*$-based radius
estimates appears considerably better, perhaps as good as 2\% RMS.
At this level, the accuracy of radius estimates is limited by
probable systematic errors not only in the evolution models,
but to a similar degree by errors in estimates of stellar $T_{\rm eff}$
and $[Z]$.
Improving this situation will require further work to obtain accurate
spectroscopic estimates of these quantities and to understand the
correlated errors between them, especially for spectra with lines that are
significantly broadened by stellar rotation.

Although the $\rho_*$ method (with YY models) works well on average, it shows
mass-dependent systematic errors of several percent when applied to components
of eclipsing binary systems.
The sense of these errors is that observed EB component radii 
and masses are larger
than those predicted by the YY models;
these differences are largest for low-mass stars.
A plausible reading of the data (but one that is not compelled by
statistics) is that the radius differences are
zero for stars with more than about 1.4 solar masses, corresponding
to the mass at which surface convection zones shrink to a negligible
fraction of the stellar radius.
For many years, it has been noted that rapidly-rotating, magnetically-active
stars may have radii that are up to 10\% larger 
than their masses and luminosities
would imply \citep{hox73,lac77,pop97}.
These claims are based for the most part on results for unequal-mass
EB components, for which it proves impossible to fit masses, radii, and
luminosities if the components are assumed to be coeval and to have
the same composition.
Examples of such stars include CV Boo B \citep{tor09} and V1061 Cyg B \citep{tor06}.
Theory \citep{chab07} suggests that the radius expansion results from blockage
of the convective energy flux by strong magnetic fields in the stellar
convection zones.
The mass dependence of EB radius 
discrepancies seen in Fig. 1 is thus likely to be
the population-averaged expression of the same physical process.

It is possible that the stellar models calculated by Baraffe \citep{chab97}
fit the observed properties of cool, low-mass stars better than the
YY models do.
These models are intended to treat such stars, using non-gray atmospheres
and smaller mixing-length parameters that act to reduce the efficiency of
energy transport in the outer stellar layers.
It would be worthwhile (but is outside the scope of this paper)
to perform a detailed comparison of the TAG masses
and radii with those inferred using the $\rho_*$ method and the
Baraffe models.

Stars with asteroseismic estimates of $\rho_*$ do not show the 
radius discrepancies found in the TAG sample.
This sample of stars are all magnetically inactive, slow rotators.
The good agreement between their observed radii and those predicted
by the YY models argues in favor of magnetic activity having a visible
influence on the radii of low-mass stars.
The number of asteroseismically-measured stars is, however, as yet too 
small to allow a firm conclusion that
these stars are drawn from a different population than the eclipsing binaries.
Future work will surely improve this situation, especially insofar as
spaceborne photometry allows asteroseismic measurements on rapidly-rotating
stars, which are difficult to measure with Doppler methods.

\acknowledgments
I am grateful to G. Torres for his valuable comments on an early draft of
this paper, and to him, J. Andersen, and A. Gim\'enez for providing
me the information on which Table 1 is based in advance of publication. 
I also thank an anonymous referee, whose suggestions materially improved
the discussion in \ss 4.2.

\clearpage

\clearpage

\begin{deluxetable}{lrrrrrrrrrrr}
\tabletypesize{\scriptsize}
\rotate
\tablecaption{Eclipsing Binary Stars with Accurate Mass and Radius
Estimates.\tablenotemark{a} \label{tbl-1}}
\tablewidth{478pt}
\tablehead{
\colhead{Star} & \colhead{$M_*$}   & \colhead{$M_*$}   &
\colhead{$R_*$} &
\colhead{$R_*$}  & \colhead{$T_{\rm eff}$} & \colhead{$T_{\rm eff}$} &
\colhead{$[Z]$\tablenotemark{b}}     & \colhead{$[Z]$\tablenotemark{b}}  &
\colhead{$M_{fit}$}   & \colhead{$R_{fit}$}           &
\colhead{Age$_{fit}$} \\
\colhead{ } & \colhead{$(M_{\sun}$)} & \colhead{RMS} & 
\colhead{($R_{\sun}$)} & 
\colhead{RMS} & \colhead{(K)} & \colhead{RMS} &
\colhead{ } & \colhead{RMS} & 
\colhead{($M_{\sun}$) } & \colhead{($R_{\sun}$)} &
\colhead{(GY)} 
}
\startdata
        U Oph B & 4.739 & 0.072 &  3.111 & 0.034 & 15590 & 250 &  0.000 & 0.200 & 4.711 &  3.104 &  0.05 \\
       DI Her B & 4.524 & 0.066 &  2.478 & 0.046 & 15100 & 700 &  0.000 & 0.200 & 4.141 &  2.406 &  0.01 \\
     V760 Sco B & 4.609 & 0.073 &  2.642 & 0.066 & 16300 & 500 &  0.000 & 0.200 & 4.757 &  2.670 &  0.01 \\
       MU Cas A & 4.657 & 0.095 &  4.196 & 0.058 & 14750 & 800 &  0.000 & 0.200 & 4.912 &  4.271 &  0.08 \\
       MU Cas B & 4.575 & 0.088 &  3.671 & 0.057 & 15100 & 800 &  0.000 & 0.200 & 4.809 &  3.732 &  0.07 \\
       GG Lup A & 4.106 & 0.044 &  2.381 & 0.025 & 14750 & 450 &  0.000 & 0.200 & 4.001 &  2.361 &  0.01 \\
       GG Lup B & 2.504 & 0.023 &  1.726 & 0.019 & 11000 & 600 &  0.000 & 0.200 & 2.414 &  1.707 &  0.04 \\
     $\zeta$ Phe A & 3.921 & 0.045 &  2.853 & 0.015 & 14400 & 800 &  0.000 & 0.200 & 4.186 &  2.915 &  0.06 \\
     $\zeta$ Phe B & 2.545 & 0.026 &  1.854 & 0.011 & 12000 & 600 &  0.000 & 0.200 & 2.867 &  1.930 &  0.03 \\
     $\chi$2 Hya A & 3.605 & 0.078 &  4.391 & 0.039 & 11750 & 190 &  0.000 & 0.200 & 3.860 &  4.492 &  0.17 \\
     $\chi$2 Hya B & 2.632 & 0.049 &  2.160 & 0.030 & 11100 & 230 &  0.000 & 0.200 & 2.780 &  2.199 &  0.14 \\
       IQ Per A & 3.504 & 0.054 &  2.446 & 0.024 & 12300 & 230 &  0.000 & 0.200 & 3.226 &  2.379 &  0.09 \\
       IQ Per B & 1.730 & 0.025 &  1.499 & 0.016 &  7700 & 140 &  0.000 & 0.200 & 1.597 &  1.460 &  0.14 \\
     V906 Sco A & 3.378 & 0.071 &  4.521 & 0.035 & 10400 & 500 &  0.140 & 0.060 & 3.490 &  4.570 &  0.23 \\
     V906 Sco B & 3.253 & 0.069 &  3.515 & 0.039 & 10700 & 500 &  0.140 & 0.060 & 3.228 &  3.506 &  0.24 \\
  OGLE 051019 A & 3.278 & 0.032 & 26.060 & 0.290 &  5300 & 100 & -0.500 & 0.100 & 4.322 & 28.577 &  0.13 \\
  OGLE 051019 B & 3.179 & 0.029 & 19.770 & 0.340 &  5450 & 100 & -0.500 & 0.100 & 3.727 & 20.846 &  0.19 \\
       PV Cas A & 2.816 & 0.050 &  2.301 & 0.020 & 10200 & 250 &  0.000 & 0.200 & 2.529 &  2.219 &  0.25 \\
       PV Cas B & 2.757 & 0.054 &  2.258 & 0.019 & 10190 & 250 &  0.000 & 0.200 & 2.514 &  2.189 &  0.24 \\
     V451 Oph A & 2.769 & 0.062 &  2.642 & 0.031 & 10800 & 800 &  0.000 & 0.200 & 2.889 &  2.679 &  0.24 \\
     V451 Oph B & 2.351 & 0.052 &  2.029 & 0.028 &  9800 & 500 &  0.000 & 0.200 & 2.347 &  2.027 &  0.24 \\
       WX Cep A & 2.533 & 0.050 &  3.997 & 0.030 &  8150 & 250 &  0.000 & 0.200 & 2.525 &  3.993 &  0.57 \\
       WX Cep B & 2.324 & 0.045 &  2.712 & 0.023 &  8900 & 250 &  0.000 & 0.200 & 2.342 &  2.719 &  0.53 \\
       TZ Men A & 2.482 & 0.025 &  2.017 & 0.020 & 10400 & 500 &  0.000 & 0.200 & 2.515 &  2.025 &  0.16 \\
       TZ Men B & 1.500 & 0.010 &  1.433 & 0.014 &  7200 & 300 &  0.000 & 0.200 & 1.494 &  1.431 &  0.18 \\
    V1031 Ori A & 2.468 & 0.018 &  4.324 & 0.034 &  7850 & 500 &  0.000 & 0.200 & 2.540 &  4.365 &  0.59 \\
    V1031 Ori B & 2.281 & 0.016 &  2.979 & 0.064 &  8400 & 500 &  0.000 & 0.200 & 2.293 &  2.984 &  0.64 \\
     V396 Cas A & 2.397 & 0.025 &  2.593 & 0.014 &  9225 & 150 &  0.000 & 0.200 & 2.388 &  2.589 &  0.46 \\
     V396 Cas B & 1.901 & 0.019 &  1.780 & 0.010 &  8550 & 120 &  0.000 & 0.200 & 1.925 &  1.787 &  0.39 \\
     $\beta$ Aur A & 2.375 & 0.027 &  2.766 & 0.018 &  9350 & 200 &  0.000 & 0.200 & 2.500 &  2.814 &  0.44 \\
     $\beta$ Aur B & 2.304 & 0.030 &  2.572 & 0.018 &  9200 & 200 &  0.000 & 0.200 & 2.386 &  2.602 &  0.47 \\
       GG Ori A & 2.342 & 0.016 &  1.854 & 0.025 &  9950 & 200 &  0.000 & 0.200 & 2.311 &  1.845 &  0.13 \\
       GG Ori B & 2.338 & 0.016 &  1.832 & 0.025 &  9950 & 200 &  0.000 & 0.200 & 2.300 &  1.822 &  0.12 \\
     V364 Lac A & 2.333 & 0.015 &  3.310 & 0.021 &  8250 & 150 &  0.000 & 0.200 & 2.355 &  3.320 &  0.64 \\
     V364 Lac B & 2.295 & 0.024 &  2.986 & 0.020 &  8500 & 150 &  0.000 & 0.200 & 2.323 &  2.998 &  0.61 \\
       YZ Cas A & 2.317 & 0.020 &  2.539 & 0.026 & 10200 & 300 &  0.000 & 0.200 & 2.697 &  2.671 &  0.31 \\
       YZ Cas B & 1.352 & 0.009 &  1.351 & 0.014 &  7200 & 300 &  0.000 & 0.200 & 1.449 &  1.385 &  0.18 \\
       SZ Cen A & 2.311 & 0.026 &  4.557 & 0.032 &  8100 & 300 &  0.000 & 0.200 & 2.785 &  4.849 &  0.47 \\
       SZ Cen B & 2.272 & 0.021 &  3.626 & 0.026 &  8380 & 300 &  0.000 & 0.200 & 2.525 &  3.755 &  0.55 \\
     V624 Her A & 2.277 & 0.014 &  3.032 & 0.051 &  8150 & 150 &  0.000 & 0.200 & 2.228 &  3.010 &  0.71 \\
     V624 Her B & 1.876 & 0.013 &  2.211 & 0.034 &  7950 & 150 &  0.000 & 0.200 & 1.923 &  2.229 &  0.83 \\
     V885 Cyg A & 2.228 & 0.026 &  3.388 & 0.026 &  8150 & 150 &  0.000 & 0.200 & 2.368 &  3.457 &  0.64 \\
     V885 Cyg B & 2.000 & 0.029 &  2.346 & 0.017 &  8375 & 150 &  0.000 & 0.200 & 2.083 &  2.378 &  0.68 \\
       GZ CMa A & 2.199 & 0.017 &  2.494 & 0.031 &  8800 & 350 &  0.000 & 0.200 & 2.241 &  2.509 &  0.56 \\
       GZ CMa B & 2.006 & 0.012 &  2.133 & 0.037 &  8500 & 350 &  0.000 & 0.200 & 2.035 &  2.143 &  0.60 \\
    V1647 Sgr A & 2.184 & 0.037 &  1.832 & 0.018 &  9600 & 300 &  0.000 & 0.200 & 2.220 &  1.842 &  0.18 \\
    V1647 Sgr B & 1.967 & 0.033 &  1.668 & 0.017 &  9100 & 300 &  0.000 & 0.200 & 2.026 &  1.684 &  0.14 \\
       EE Peg A & 2.151 & 0.024 &  2.090 & 0.025 &  8700 & 200 &  0.000 & 0.200 & 2.064 &  2.061 &  0.51 \\
       EE Peg B & 1.332 & 0.011 &  1.312 & 0.013 &  6450 & 300 &  0.000 & 0.200 & 1.340 &  1.314 &  0.46 \\
       AI Hya A & 2.140 & 0.038 &  3.917 & 0.031 &  6700 &  60 &  0.000 & 0.200 & 2.072 &  3.875 &  1.06 \\
       AI Hya B & 1.973 & 0.036 &  2.768 & 0.019 &  7100 &  65 &  0.000 & 0.200 & 1.854 &  2.711 &  1.21 \\
       VV Pyx A & 2.097 & 0.022 &  2.169 & 0.020 &  9500 & 200 &  0.000 & 0.200 & 2.347 &  2.251 &  0.37 \\
       VV Pyx B & 2.095 & 0.019 &  2.169 & 0.020 &  9500 & 200 &  0.000 & 0.200 & 2.347 &  2.252 &  0.37 \\
       TZ For A & 2.045 & 0.055 &  8.330 & 0.120 &  5000 & 100 &  0.100 & 0.150 & 2.489 &  8.893 &  0.66 \\
       TZ For B & 1.945 & 0.027 &  3.966 & 0.088 &  6350 & 100 &  0.100 & 0.150 & 2.135 &  4.087 &  1.05 \\
     V459 Cas A & 2.030 & 0.036 &  2.014 & 0.020 &  9140 & 300 &  0.000 & 0.200 & 2.175 &  2.061 &  0.39 \\
     V459 Cas B & 1.973 & 0.034 &  1.970 & 0.020 &  9100 & 300 &  0.000 & 0.200 & 2.153 &  2.028 &  0.39 \\
       EK Cep A & 2.025 & 0.023 &  1.580 & 0.007 &  9000 & 200 &  0.070 & 0.050 & 1.928 &  1.583 &  0.08 \\
       EK Cep B & 1.122 & 0.012 &  1.316 & 0.006 &  5700 & 200 &  0.070 & 0.050 & 1.042 &  1.283 &  7.71 \\
       KW Hya A & 1.973 & 0.036 &  2.127 & 0.016 &  8000 & 200 &  0.000 & 0.200 & 1.893 &  2.098 &  0.78 \\
       KW Hya B & 1.485 & 0.017 &  1.480 & 0.013 &  6900 & 200 &  0.000 & 0.200 & 1.450 &  1.468 &  0.60 \\
       WW Aur A & 1.964 & 0.010 &  1.929 & 0.011 &  7960 & 420 &  0.000 & 0.200 & 1.813 &  1.878 &  0.69 \\
       WW Aur B & 1.814 & 0.008 &  1.838 & 0.011 &  7670 & 410 &  0.000 & 0.200 & 1.714 &  1.803 &  0.79 \\
       WW Cam A & 1.920 & 0.013 &  1.912 & 0.016 &  8350 & 135 &  0.000 & 0.200 & 1.917 &  1.911 &  0.56 \\
       WW Cam B & 1.873 & 0.018 &  1.809 & 0.016 &  8240 & 135 &  0.000 & 0.200 & 1.852 &  1.802 &  0.52 \\
     V392 Car A & 1.904 & 0.013 &  1.625 & 0.022 &  8850 & 200 &  0.000 & 0.200 & 1.945 &  1.636 &  0.14 \\
     V392 Car B & 1.855 & 0.020 &  1.601 & 0.022 &  8630 & 200 &  0.000 & 0.200 & 1.884 &  1.609 &  0.14 \\
       AY Cam A & 1.901 & 0.040 &  2.771 & 0.020 &  7250 & 100 &  0.000 & 0.200 & 1.914 &  2.777 &  1.11 \\
       AY Cam B & 1.706 & 0.036 &  2.025 & 0.017 &  7395 & 100 &  0.000 & 0.200 & 1.715 &  2.028 &  1.10 \\
       RS Cha A & 1.854 & 0.016 &  2.139 & 0.055 &  8050 & 200 &  0.170 & 0.010 & 2.010 &  2.197 &  0.63 \\
       RS Cha B & 1.817 & 0.018 &  2.340 & 0.055 &  7700 & 200 &  0.170 & 0.010 & 1.986 &  2.410 &  0.81 \\
       MY Cyg A & 1.806 & 0.025 &  2.243 & 0.050 &  7050 & 200 &  0.000 & 0.200 & 1.683 &  2.191 &  1.36 \\
       MY Cyg B & 1.782 & 0.030 &  2.178 & 0.050 &  7000 & 200 &  0.000 & 0.200 & 1.648 &  2.122 &  1.42 \\
       EI Cep A & 1.772 & 0.007 &  2.898 & 0.048 &  6750 & 100 &  0.000 & 0.200 & 1.828 &  2.928 &  1.36 \\
       EI Cep B & 1.680 & 0.006 &  2.331 & 0.044 &  6950 & 100 &  0.000 & 0.200 & 1.703 &  2.341 &  1.42 \\
       BP Vul A & 1.737 & 0.015 &  1.853 & 0.014 &  7715 & 150 &  0.000 & 0.200 & 1.738 &  1.853 &  0.81 \\
       BP Vul B & 1.408 & 0.009 &  1.490 & 0.013 &  6810 & 150 &  0.000 & 0.200 & 1.423 &  1.495 &  1.00 \\
       FS Mon A & 1.632 & 0.010 &  2.052 & 0.012 &  6715 & 100 &  0.000 & 0.200 & 1.539 &  2.012 &  1.82 \\
       FS Mon B & 1.462 & 0.009 &  1.630 & 0.010 &  6550 & 100 &  0.000 & 0.200 & 1.348 &  1.586 &  2.28 \\
       PV Pup A & 1.561 & 0.011 &  1.544 & 0.016 &  6920 & 300 &  0.000 & 0.200 & 1.449 &  1.506 &  0.86 \\
       PV Pup B & 1.550 & 0.013 &  1.500 & 0.016 &  6930 & 300 &  0.000 & 0.200 & 1.468 &  1.472 &  0.50 \\
     V442 Cyg A & 1.560 & 0.024 &  2.074 & 0.034 &  6900 & 100 &  0.000 & 0.200 & 1.615 &  2.098 &  1.53 \\
     V442 Cyg B & 1.407 & 0.023 &  1.663 & 0.033 &  6800 & 100 &  0.000 & 0.200 & 1.467 &  1.686 &  1.52 \\
       EY Cep A & 1.520 & 0.012 &  1.464 & 0.011 &  7090 & 150 &  0.000 & 0.200 & 1.494 &  1.455 &  0.25 \\
       EY Cep B & 1.496 & 0.016 &  1.471 & 0.011 &  6970 & 150 &  0.000 & 0.200 & 1.470 &  1.462 &  0.42 \\
     HD 71636 A & 1.511 & 0.007 &  1.570 & 0.026 &  6950 & 140 &  0.000 & 0.200 & 1.467 &  1.555 &  0.98 \\
     HD 71636 B & 1.285 & 0.006 &  1.362 & 0.026 &  6440 & 140 &  0.000 & 0.200 & 1.283 &  1.361 &  1.83 \\
       RZ Cha A & 1.493 & 0.022 &  2.257 & 0.016 &  6450 & 150 &  0.000 & 0.200 & 1.556 &  2.288 &  2.11 \\
       RZ Cha B & 1.493 & 0.022 &  2.257 & 0.016 &  6450 & 150 &  0.000 & 0.200 & 1.399 &  2.209 &  3.03 \\
       GX Gem A & 1.488 & 0.011 &  2.327 & 0.012 &  6195 & 100 &  0.000 & 0.200 & 1.446 &  2.305 &  3.09 \\
       GX Gem B & 1.467 & 0.010 &  2.236 & 0.012 &  6165 & 100 &  0.000 & 0.200 & 1.484 &  2.245 &  2.69 \\
       BW Aqr A & 1.479 & 0.019 &  2.063 & 0.044 &  6350 & 100 &  0.000 & 0.200 & 1.463 &  2.056 &  2.50 \\
       BW Aqr B & 1.377 & 0.021 &  1.786 & 0.043 &  6450 & 100 &  0.000 & 0.200 & 1.411 &  1.801 &  2.41 \\
       DM Vir A & 1.454 & 0.008 &  1.765 & 0.017 &  6500 & 100 &  0.000 & 0.200 & 1.402 &  1.744 &  2.31 \\
       DM Vir B & 1.448 & 0.008 &  1.765 & 0.017 &  6500 & 300 &  0.000 & 0.200 & 1.424 &  1.755 &  2.12 \\
     V570 Per A & 1.447 & 0.009 &  1.521 & 0.034 &  6842 &  50 &  0.020 & 0.030 & 1.423 &  1.512 &  1.10 \\
     V570 Per B & 1.347 & 0.008 &  1.386 & 0.019 &  6562 &  50 &  0.020 & 0.030 & 1.308 &  1.372 &  1.48 \\
       CD Tau A & 1.442 & 0.016 &  1.798 & 0.015 &  6200 &  50 &  0.080 & 0.150 & 1.358 &  1.762 &  3.11 \\
       CD Tau B & 1.368 & 0.016 &  1.585 & 0.018 &  6200 &  50 &  0.080 & 0.150 & 1.289 &  1.554 &  3.23 \\
       AD Boo A & 1.414 & 0.009 &  1.614 & 0.014 &  6575 & 120 &  0.100 & 0.150 & 1.418 &  1.615 &  1.75 \\
       AD Boo B & 1.209 & 0.006 &  1.217 & 0.010 &  6145 & 120 &  0.100 & 0.150 & 1.200 &  1.214 &  1.97 \\
    V1143 Cyg A & 1.388 & 0.016 &  1.347 & 0.023 &  6450 & 100 &  0.000 & 0.200 & 1.269 &  1.307 &  1.50 \\
    V1143 Cyg B & 1.344 & 0.013 &  1.324 & 0.023 &  6400 & 100 &  0.000 & 0.200 & 1.257 &  1.295 &  1.60 \\
       IT Cas A & 1.332 & 0.009 &  1.595 & 0.018 &  6470 & 100 &  0.000 & 0.200 & 1.349 &  1.601 &  2.42 \\
       IT Cas B & 1.329 & 0.008 &  1.563 & 0.018 &  6470 & 100 &  0.000 & 0.200 & 1.333 &  1.564 &  2.43 \\
    V1061 Cyg A & 1.282 & 0.015 &  1.616 & 0.017 &  6180 & 100 &  0.000 & 0.200 & 1.287 &  1.618 &  3.46 \\
    V1061 Cyg B & 0.932 & 0.007 &  0.967 & 0.011 &  5300 & 150 &  0.000 & 0.200 & 0.809 &  0.923 & 16.36 \\
       VZ Hya A & 1.271 & 0.009 &  1.314 & 0.005 &  6645 & 150 & -0.200 & 0.120 & 1.258 &  1.309 &  1.41 \\
       VZ Hya B & 1.146 & 0.006 &  1.113 & 0.007 &  6290 & 150 & -0.200 & 0.120 & 1.124 &  1.105 &  1.51 \\
     V505 Per A & 1.269 & 0.007 &  1.287 & 0.024 &  6510 &  50 & -0.120 & 0.030 & 1.222 &  1.270 &  1.71 \\
     V505 Per B & 1.251 & 0.007 &  1.266 & 0.024 &  6460 &  50 & -0.120 & 0.030 & 1.201 &  1.249 &  1.87 \\
       HS Hya A & 1.255 & 0.008 &  1.276 & 0.007 &  6500 &  50 &  0.000 & 0.200 & 1.297 &  1.289 &  0.87 \\
       HS Hya B & 1.219 & 0.007 &  1.217 & 0.007 &  6400 &  50 &  0.000 & 0.200 & 1.259 &  1.229 &  0.78 \\
       RT And A & 1.240 & 0.030 &  1.256 & 0.015 &  6100 & 150 &  0.000 & 0.200 & 1.154 &  1.226 &  3.07 \\
       RT And B & 0.907 & 0.017 &  0.907 & 0.011 &  4880 & 100 &  0.000 & 0.200 & 0.727 &  0.840 & 25.68 \\
       UX Men A & 1.235 & 0.006 &  1.348 & 0.013 &  6200 & 100 &  0.040 & 0.100 & 1.205 &  1.337 &  3.05 \\
       UX Men B & 1.196 & 0.007 &  1.275 & 0.013 &  6150 & 100 &  0.040 & 0.100 & 1.173 &  1.267 &  3.05 \\
       AI Phe A & 1.234 & 0.004 &  2.932 & 0.048 &  5010 & 120 & -0.140 & 0.100 & 1.292 &  2.978 &  4.24 \\
       AI Phe B & 1.193 & 0.004 &  1.818 & 0.024 &  6310 & 150 & -0.140 & 0.100 & 1.360 &  1.899 &  2.92 \\
       WZ Oph A & 1.227 & 0.007 &  1.402 & 0.012 &  6165 & 100 & -0.270 & 0.070 & 1.045 &  1.331 &  5.98 \\
       WZ Oph B & 1.220 & 0.006 &  1.420 & 0.012 &  6115 & 100 & -0.270 & 0.070 & 1.044 &  1.347 &  6.17 \\
       FL Lyr A & 1.218 & 0.016 &  1.283 & 0.028 &  6150 & 100 &  0.000 & 0.200 & 1.173 &  1.267 &  3.05 \\
       FL Lyr B & 0.958 & 0.012 &  0.962 & 0.028 &  5300 & 100 &  0.000 & 0.200 & 0.815 &  0.913 & 15.25 \\
     V432 Aur A & 1.204 & 0.006 &  2.430 & 0.023 &  6080 &  85 & -0.600 & 0.050 & 1.168 &  2.405 &  4.36 \\
     V432 Aur B & 1.079 & 0.005 &  1.224 & 0.007 &  6685 &  85 & -0.600 & 0.050 & 1.071 &  1.220 &  3.55 \\
       EW Ori A & 1.174 & 0.012 &  1.134 & 0.011 &  5970 & 100 &  0.000 & 0.200 & 1.074 &  1.100 &  3.67 \\
       EW Ori B & 1.124 & 0.009 &  1.083 & 0.011 &  5780 & 100 &  0.000 & 0.200 & 0.989 &  1.038 &  6.03 \\
       BH Vir A & 1.166 & 0.008 &  1.247 & 0.024 &  6100 & 100 &  0.000 & 0.200 & 1.162 &  1.246 &  3.12 \\
       BH Vir B & 1.052 & 0.006 &  1.135 & 0.023 &  5500 & 200 &  0.000 & 0.200 & 0.929 &  1.086 & 11.04 \\
       ZZ UMa A & 1.139 & 0.005 &  1.514 & 0.019 &  5960 &  70 &  0.000 & 0.200 & 1.148 &  1.516 &  5.54 \\
       ZZ UMa B & 0.969 & 0.005 &  1.156 & 0.010 &  5270 &  90 &  0.000 & 0.200 & 0.846 &  1.103 & 18.47 \\
    $\alpha$ Cen A & 1.105 & 0.007 &  1.224 & 0.003 &  5824 &  26 &  0.240 & 0.040 & 1.124 &  1.231 &  4.69 \\
    $\alpha$ Cen B & 0.934 & 0.006 &  0.863 & 0.005 &  5223 &  62 &  0.240 & 0.040 & 0.926 &  0.859 &  3.79 \\
NGC188 KR V12 A & 1.103 & 0.007 &  1.426 & 0.019 &  5900 & 100 & -0.100 & 0.090 & 1.043 &  1.401 &  7.57 \\
NGC188 KR V12 B & 1.081 & 0.007 &  1.374 & 0.019 &  5875 & 100 & -0.100 & 0.090 & 1.052 &  1.358 &  7.03 \\
     V568 Lyr A & 1.074 & 0.008 &  1.400 & 0.016 &  5665 & 100 &  0.400 & 0.100 & 1.200 &  1.452 &  4.91 \\
     V568 Lyr B & 0.827 & 0.004 &  0.768 & 0.006 &  4900 & 100 &  0.400 & 0.100 & 0.861 &  0.779 &  3.04 \\
     V636 Cen A & 1.052 & 0.005 &  1.019 & 0.004 &  5900 &  85 & -0.200 & 0.080 & 0.958 &  0.988 &  5.42 \\
     V636 Cen B & 0.854 & 0.003 &  0.830 & 0.004 &  5000 & 100 & -0.200 & 0.080 & 0.681 &  0.771 & 25.55 \\
       CV Boo A & 1.032 & 0.013 &  1.263 & 0.023 &  5760 & 150 &  0.000 & 0.200 & 1.029 &  1.262 &  7.91 \\
       CV Boo B & 0.968 & 0.012 &  1.174 & 0.023 &  5670 & 150 &  0.000 & 0.200 & 0.925 &  1.160 & 11.80 \\
    V1174 Ori A & 1.006 & 0.013 &  1.338 & 0.011 &  4470 & 120 &  0.000 & 0.200 & 0.698 &  1.184 & 42.39 \\
    V1174 Ori B & 0.727 & 0.010 &  1.063 & 0.011 &  3615 & 100 &  0.000 & 0.200 & 0.428 &  0.891 & 108.00 \\
       UV Psc A & 0.983 & 0.008 &  1.110 & 0.023 &  5780 & 100 &  0.000 & 0.200 & 0.981 &  1.108 &  7.79 \\
       UV Psc B & 0.764 & 0.004 &  0.835 & 0.018 &  4750 &  80 &  0.000 & 0.200 & 0.710 &  0.813 & 28.42 \\
       CG Cyg A & 0.941 & 0.014 &  0.893 & 0.012 &  5260 & 180 &  0.000 & 0.200 & 0.834 &  0.858 & 11.18 \\
       CG Cyg B & 0.815 & 0.013 &  0.838 & 0.011 &  4720 &  60 &  0.000 & 0.200 & 0.700 &  0.795 & 29.00 \\
       RW Lac A & 0.926 & 0.006 &  1.187 & 0.004 &  5760 & 100 &  0.000 & 0.200 & 0.973 &  1.210 &  9.63 \\
       RW Lac B & 0.869 & 0.004 &  0.964 & 0.004 &  5560 & 150 &  0.000 & 0.200 & 0.905 &  0.977 &  9.64 \\
       HS Aur A & 0.898 & 0.019 &  1.005 & 0.024 &  5350 &  75 &  0.000 & 0.200 & 0.834 &  0.980 & 15.82 \\
       HS Aur B & 0.877 & 0.017 &  0.874 & 0.024 &  5200 &  75 &  0.000 & 0.200 & 0.813 &  0.852 & 13.09 \\
       GU Boo A & 0.609 & 0.006 &  0.627 & 0.016 &  3920 & 130 &  0.000 & 0.200 & 0.514 &  0.592 & 56.22 \\
       GU Boo B & 0.598 & 0.006 &  0.623 & 0.016 &  3810 & 130 &  0.000 & 0.200 & 0.493 &  0.584 & 61.11 \\
       YY Gem A & 0.599 & 0.005 &  0.619 & 0.006 &  3820 & 100 &  0.000 & 0.200 & 0.498 &  0.581 & 57.80 \\
       YY Gem B & 0.599 & 0.005 &  0.619 & 0.006 &  3820 & 100 &  0.000 & 0.200 & 0.498 &  0.581 & 57.81 \\
       CU Cnc A & 0.435 & 0.001 &  0.432 & 0.005 &  3160 & 150 &  0.000 & 0.200 & 0.425 &  0.434 &  1.40 \\
\enddata
\tablenotetext{a}{Data in all columns except the last 3 and in some cases $[Z]$ and
$[Z]_{RMS}$ (as detailed below) are from \citet{tor09} (TAG)}
\tablenotetext{b}{For stars with no observational estimate of $[Z]$, I
have set $[Z] = 0.$, and $[Z]_{RMS} = 0.2$.}
\end{deluxetable}

\clearpage

\begin{deluxetable}{lrrrrrrrrrrrr}
\tabletypesize{\scriptsize}
\rotate
\tablecaption{Stars with $M_*$ and $R_*$ estimates based on 
asteroseismology. \label{tbl-2}}
\tablewidth{0pt}
\tablehead{
\colhead{Star} & \colhead{$M_*$}   & \colhead{$M_*$}   &
\colhead{$R_*$} &
\colhead{$R_*$}  & \colhead{$T_{\rm eff}$} & \colhead{$T_{\rm eff}$} &
\colhead{$[Z]$}     & \colhead{$[Z]$}  &
\colhead{$M_{fit}$}   & \colhead{$R_{fit}$}           &
\colhead{Age$_{fit}$} & \colhead{Source} \\
\colhead{ } & \colhead{$(M_{\sun}$)} & \colhead{RMS} & 
\colhead{($R_{\sun}$)} & 
\colhead{RMS} & \colhead{(K)} & \colhead{RMS} &
\colhead{ } & \colhead{RMS} & 
\colhead{($M_{\sun}$) } & \colhead{($R_{\sun}$)} &
\colhead{(GY)} & \colhead{ }
}
\startdata

        HD49933 & 1.079 & 0.073 &  1.385 & 0.031 &  6650 &  75 & -0.440 & 0.060 & 1.150 &  1.414 &  3.47 &  \citet{bru09} \\
       HD175726 & 0.993 & 0.060 &  1.014 & 0.035 &  6060 &  50 & -0.100 & 0.060 & 1.063 &  1.037 &  2.34 &  \citet{bru09} \\
       HD181420 & 1.311 & 0.063 &  1.595 & 0.032 &  6620 & 100 &  0.000 & 0.120 & 1.391 &  1.627 &  1.96 &  \citet{bru09} \\
       HD181906 & 1.144 & 0.119 &  1.392 & 0.054 &  6365 & 120 & -0.110 & 0.140 & 1.224 &  1.423 &  3.10 &  \citet{bru09} \\
       70 Oph A & 0.895 & 0.005 &  0.863 & 0.002 &  5322 &  40 &  0.011 & 0.050 & 0.870 &  0.855 &  7.56 &  \citet{tan08} \\
    $\alpha$ Cen A & 1.105 & 0.007 &  1.224 & 0.003 &  5824 &  26 &  0.240 & 0.040 & 1.124 &  1.231 &  4.69 &  \citet{tor09} \\
    $\alpha$ Cen B & 0.934 & 0.006 &  0.863 & 0.005 &  5223 &  62 &  0.240 & 0.040 & 0.926 &  0.859 &  3.79 &  \citet{tor09} \\
         $\mu$ Ara & 1.100 & 0.010 &  1.353 & 0.010 &  5800 &  90 &  0.300 & 0.050 & 1.190 &  1.389 &  4.54 &  \citet{sor09} \\
        Procyon & 1.497 & 0.037 &  2.067 & 0.028 &  6530 &  90 & -0.050 & 0.030 & 1.477 &  2.058 &  2.24 &  \citet{egg05} \\
        $\eta$ Boo & 1.700 & 0.050 &  2.790 & 0.040 &  6030 &  90 &  0.360 & 0.050 & 1.715 &  2.798 &  1.96 &  \citet{car05} \\
       $\beta$ Hyi & 1.070 & 0.030 &  1.814 & 0.017 &  5872 &  44 & -0.030 & 0.050 & 1.168 &  1.867 &  5.82 &  \citet{nor07} \\
      $\delta$ Eri & 1.215 & 0.020 &  2.328 & 0.010 &  5074 &  60 &  0.130 & 0.080 & 1.234 &  2.344 &  5.61 &  \citet{the05} \\
      $\zeta$ Hyi & 2.650 & 0.020 & 10.300 & 0.100 &  5010 & 100 & -0.040 & 0.120 & 2.777 & 10.462 &  0.47 &  \citet{the05} \\
      $\beta$ Vir & 1.413 & 0.061 &  1.703 & 0.022 &  6059 &  49 &  0.140 & 0.050 & 1.301 &  1.656 &  3.65 & \citet{nor09} \\
       $\tau$ Cet & 0.783 & 0.012 &  0.793 & 0.004 &  5265 & 100 & -0.500 & 0.030 & 0.660 &  0.751 & 24.29 & \citet{tei09} \\
\enddata
\end{deluxetable}

\end{document}